\documentclass[a4paper,11pt]{article}
\usepackage{pos}
\usepackage{amsmath,graphicx,multirow,tabularx,physics}
\usepackage{davitex}

\title{Testing strong isospin breaking effects in QCD thermodynamics}

\author*[a]{D. A. Clarke}
\author[a]{B. B. Brandt}

\affiliation[a]{Fakult\"at f\"ur Physik, Universit\"at Bielefeld, \\
D-33615 Bielefeld, Germany}

\emailAdd{clarke.davida@gmail.com}

\abstract{
Most modern lattice QCD calculations take the masses of the light quarks to be
degenerate, setting $m_u=m_d$, a choice that, while not physical, increases
computational efficiency and lowers the complexity of simulations and analysis.
For many observables, the systematic error associated with this simplification
is expected to be negligible, as $\Delta m_{ud}=m_d-m_u$ 
is significantly below the QCD scale. So far
lattice studies have examined the role of strong isospin breaking (SIB) in
hadron mass splittings, but SIB effects in thermodynamic observables have only
been studied in terms of the crossover temperature with unphysically heavy
quarks. Motivated by the recent observation of an unexpectedly large ratio
between the abundances of charged versus neutral kaons in heavy-ion collision
experiments, we investigate the effects of SIB in QCD thermodynamics by computing
observables for three non-degenerate flavors ($N_f=1+1+1$) with physical quark masses using dynamical improved staggered fermions. In particular, we investigate the difference between the $u$- and $d$-quark condensates and extract the pseudocritical temperature, the equation
of state, and conserved-charge fluctuations, comparing results to standard
$N_f=2+1$ simulations.}

\FullConference{The 43rd International Symposium on Lattice Field Theory (Lattice 2026)\\
July 26 to August 1, 2026\\
University of Maryland, College Park, USA\\}

\newcommand{\pbpq}{\ev{\bar{q}q}}

\newcommand{\pbps}{\ev{\bar{s}s}}

\begin{document}
\maketitle

\section{Motivation}

Effects due to strong isospin breaking (SIB), i.e. effects due to the
small difference between the up-quark mass $m_u$ and down-quark mass $m_d$,
are expected to be small, as $\Delta m_{ud}\equiv m_d-m_u$ lies significantly below
the QCD scale $\Lambda_{\rm QCD}$. This manifests, for instance, in the parametrically small difference between the masses of hadrons differing only in their isospin quantum number~\cite{ParticleDataGroup:2026mpi}.
Thus lattice simulations mostly employ the isospin-symmetric
limit, where sea-quark masses are taken to be degenerate, $m_u=m_d\equiv m_l$.
In recent years lattice computations of standard model and hadron physics observables are approaching sub-percent precision, see the recent FLAG review~\cite{FlavourLatticeAveragingGroupFLAG:2024oxs}, for instance, necessitating the inclusion of SIB effects. In particular, SIB effects on the hadron spectrum have been investigated and show good agreement with experimentally observed splittings, e.g.~
\cite{deDivitiis:2011eh,BMW:2014pzb,Giusti:2017dmp,Frezzotti:2022dwn}, leading to the conclusion that SIB effects are indeed suppressed in hadronic quantities.

Similarly, SIB effects in QCD thermodynamics in the vicinity of the phase change to the quark gluon plasma can be expected to play a subleading role, since $\Delta m_{ud}$ is smaller than the relevant energy scale, the crossover temperature $\Tpc\approx 157$~MeV~\cite{HotQCD:2018pds,Borsanyi:2020fev}, so that studies in the literature are mainly limited to the isospin-symmetric limit, see~\cite{Brandt:2025sut} for a recent review.
The only study of SIB effects in the literature, Ref.~\cite{Gavai:2002fi}, reports no discernible impact of SIB, but is limited to a study of the effects on $\Tpc$ on one coarse lattice spacing and large unphysical pion masses. However, while SIB effects in hadrons can be suppressed further through the mass generation due to the strong gluon fields, it remains to be seen whether these effects remain small in all thermodynamic observables when quarks are gradually liberated with increasing temperatures and densities.
While the effect of SIB through quark masses on QCD thermodynamics has so far not been investigated systematically, the explicit breaking of isospin symmetry through the presence of external magnetic fields and differences in quark chemical potentials, the presence of an isospin chemical potential, has been studied extensively and shown to lead to novel phenomena, see Ref.~\cite{Brandt:2026dcd} for a recent review.

The question concerning possible SIB effects in QCD thermodynamics also has recently received a new motivation by the observation of an unexpectedly large difference in the abundances of charged and neutral kaons by NA61/SHINE~\cite{NA61SHINE:2023azp} in high-energy nucleus-nucleus collisions for a large region of collision energy. The average of the abundance ratio in this energy regime leads to (see Refs.~\cite{NA61SHINE:2023azp,Gazdzicki:2026bby} for discussions and details)
\begin{equation}
    R_K \equiv \frac{\ev{K^+}+\ev{K^-}}{\ev{K^0}+\ev{\bar{K}^0}} \approx 1.18 \,.
\end{equation}
This is a surprising result: 
In heavy-ion experiments, the collided nuclei
typically have more neutrons than protons, implying an excess of $d$ quarks in the final state and hence $R_K<1$.
Model calculations for the hadron abundances, such as statistical hadronization or UrQMD,
predict an excess of charged over neutral kaons, but lead to a ratio of $R_K\lesssim 1.05$ (see~\cite{NA61SHINE:2023azp}) and cannot explain the observed discrepancy.

In Ref.~\cite{Kapusta:2024zmd} it has been proposed that anomalous correlations in charged and neutral kaon fluctuations might be explained by collective regions with non-zero isospin components of the chiral condensate, dubbed disoriented chiral condensates. As argued there, the presence of the resulting difference between $\ev{\bar{u}u}$ and~$\ev{\bar{d}d}$ condensates might lead to different abundances of $\bar{u}u$ vs. $\bar{d}d$ pairs and, consequently, charged and neutral kaons. In the presence of global SIB through $\Delta m_{ud}$, the $I=1$, $I_3=0$ component of the chiral condensate is not forced to vanish from regular symmetry arguments, due to the explicit breaking of SU$_V$(2) symmetry, potentially leading to $\ev{\bar{u}u}\neq\ev{\bar{d}d}$ globally. Following the same argument as in Ref.~\cite{Kapusta:2024zmd}, this can potentially also affect the abundance ratio of charged and neutral kaons. While the mechanism needs to be understood better 
phenomenologically,
observing a difference between $\ev{\bar{u}u}$ and $\ev{\bar{d}d}$ at physical quark mass splitting might open up a new direction for explaining the observed ratio of kaon abundances.

Motivated both by the missing crosscheck of the size of SIB effects in QCD thermodynamics as well as the experimentally observed charged kaon abundances, we carry out the first study of SIB effects on QCD thermodynamics with improved staggered quarks at the physical point, albeit at a single lattice spacing. In these proceedings we report on the setup and current status of results. Results with updated statistics and extended discussions will be given in a future publication.

\section{Observables}

The observables tied to the chiral crossover are the chiral condensates, $\ev{\bar u u}$ and $\ev{\bar d d}$, which need multiplicative and additive renormalization. Since the additive renormalization is temperature independent, additive renormalization can be taken care of by using the subtracted condensate
$\Delta_q\equiv\pbpq_T-\pbpq_0$,
where the subscript indicates the temperature at which the
condensate is evaluated, and the difference is to be computed
at fixed lattice spacing.
Multiplicative renormalization can be taken care of through multiplication with the quark mass.
We thus follow Refs.~\cite{Bali:2012zg,Brandt:2017oyy} and use the fully renormalized chiral condensates
\begin{equation}\label{eq:ccren}
    \Sigma_{\bar q q}=\frac{2m_l\Delta_q}{m_\pi^2f_\pi^2}+1,
\end{equation}
with $m_l\equiv(m_u+m_d)/2$.
The normalization ensures $\Sigma_{\bar q q}\to1$ for $T=0$ and $\Sigma_{\bar q q}\to0$ for $T\to\infty$, following GMOR.
Crossover temperatures can be defined through the inflection points of the chiral condensates. While this can be done for individual condensates, the chiral crossover temperature $\Tpc$ is defined via the inflection point of the iso-singlet condensate
\begin{equation}
    \Sigma_{ud}=\frac{1}{2}\left(\Sigma_{\bar u u}+\Sigma_{\bar d d}\right).
\end{equation}

To address the magnitude of the scalar iso-vector condensate, we would like to compute the difference between the condensates normalized to the iso-singlet one.
Using the GMOR relation and Eq.~\eqref{eq:ccren}, the iso-singlet condensate
and the additively renormalized condensates $\Delta_q$
can be used to define the renormalized, normalized difference (abundance)
\begin{equation}\label{eq:A1}
    A_1\equiv\frac{m_l\left(\Delta_d-\Delta_u\right)}{
    m_l\left(\Delta_d+\Delta_u\right)+m_\pi^2f_\pi^2}.
\end{equation}
The difference tends to 0 as $T\to0$,
which might be incorrect in case of a non-zero scalar iso-vector component
in this limit.
Another possibility to define a partly renormalized difference is
to use the strange-quark condensate subtraction,
$\delta_q\equiv\pbpq-\frac{m_q}{m_s}\pbps$,
following Ref.~\cite{Cheng:2007jq}.
This eliminates the leading additive divergence,
but leaves a logarithmic divergence remaining.
Then an alternative normalized difference can be defined as
\begin{equation}\label{eq:A3}
    A_2\equiv\frac{\delta_d-\delta_u}{\delta_d+\delta_u},
\end{equation}
which we use to crosscheck the results obtained from $A_1$.

We are also interested in SIB effects on bulk thermodynamics,
the equation of state in particular.
To this end we are computing the QCD pressure $\hat p\equiv p/T^4$,
from which many other thermodynamic observables can be derived, and second-order
cumulants for the continuation to non-zero chemical potentials.
We obtain the pressure from the trace anomaly $\Theta^{\mu\mu}$
using the integral method,
\begin{equation}
    \frac{P(T)}{T^4}-\frac{P(T_0)}{T_0^4}
=\int_{T_0}^T \dd{T'}\frac{\Theta^{\mu\mu}(T')}{T'^5},
\end{equation}
where the integration is performed starting at some reference temperature $T_0$.
To compute $\Theta^{\mu\mu}$ from the lattice, we
express it as a derivative of the partition function with respect to the lattice scale,
as pioneered in Ref.~\cite{Engels:1990vr} (see also Refs.~\cite{Borsanyi:2010cj,HotQCD:2014kol} for more details).

The QCD pressure at finite $\vec\mu$ can be expanded~\cite{Allton:2002zi} in terms of the
baryon-number, electric-charge, and strangeness
chemical potentials $\hat\mu_B$, $\hat\mu_Q$, $\hat\mu_S$ (in units of $T$) as
\begin{equation}
\hat p\left(T,\vec\mu\right) = \sum_{i,j,k=0}^\infty
\frac{\chi_{ijk}^{BQS}}{i!j!k!} \hat\mu_B^i \hat\mu_Q^j \hat\mu_S^k \; ,
\end{equation}
defining the generalized susceptibilities
$\chi_{ijk}^{BQS}$ in the
$(B,Q,S)$ basis. Besides their use for reconstructing the pressure series,
these cumulants can be employed to estimate its convergence radius,
and certain cumulant ratios contain information about phase transitions.
Of special interest for SIB studies is the $(B,I,S)$ basis
with isospin $I$ (see~\cite{Brandt:2025sut}).
In particular, $\chi_{11}^{BI}(T)$ is identically
zero in the limit of exact isospin symmetry.

\section{Computational set up}

Our simulations have been done using $N_f=1+1+1$ flavors of dynamical staggered fermions
with a two-times stout-smeared action and a tree-level-improved Symanzik
gauge action, matching the setting already employed in Ref.~\cite{Borsanyi:2010cj}.
The simulation code has already been used extensively at non-zero
external magnetic fields, e.g., Ref.~\cite{Bali:2011qj}, and non-zero isospin chemical
potential, e.g., Ref.~\cite{Brandt:2017oyy}.
The approximate bare quark mass ratios
$m_u:m_d:m_s$ are
$1:2:41$. To be more precise, from the isospin symmetric estimate of $m_s$ we obtain the isospin asymmetric light quark masses via
$m_d = m_s/20.1$ and $m_u=0.49 \,m_d$.
The ratios are taken from the $N_f=2+1$ values from the latest FLAG review~\cite{FlavourLatticeAveragingGroupFLAG:2024oxs}.
The strange quark mass follows from the line of constant physics determined in
Ref.~\cite{Borsanyi:2010cj}.
The physical value of the pion mass is given by $m_\pi=135$~MeV,
according to the neutral pion mass.
For determining the condensates used for additive renormalization
and for checking the lattice spacing and hadron masses that we obtain
after tuning the light quark masses,
we perform simulations at vanishing temperature for four different
lattice spacings. On these ensembles we have explicitly checked the agreement with
the pion mass and the lattice spacing from the line of constant physics
of Ref.~\cite{Borsanyi:2010cj} and that we obtain a physical mass splitting
between charged and neutral kaons. 
For our study of thermodynamics, temperatures run from about 114 MeV to about 174 MeV
on lattices of size $24^3\times 8$. To get a quantitative handle on the impact
of non-degenerate light quark masses,
we supplement our data set with $N_f=2+1$ configurations for $T>0$
with $m_u=m_d\equiv m_l$ and mass ratio $m_l:m_s$ of $1:28.15$.
For the pion decay constant, we use a $\chi_{\rm PT}$
result $f_\pi=86.2$ MeV~\cite{Colangelo:2003hf}. 
This leads to
$m_\pi^2 f_\pi^2 = 0.08929$ $[1/{\rm fm}^4]$.

In plots, we use $\Tpc=157.3(2.3)$ MeV, an average between
Refs.~\cite{HotQCD:2018pds} and \cite{Borsanyi:2020fev}.
For the pressure, we take $T_0=100$ MeV as a reference temperature
using the hadron resonance gas (HRG) model to supply reference
values for $\hat p$ and $\hat\Theta^{\mu\mu}$.
Data analysis is carried out using the software of the
AnalysisToolbox~\cite{Clarke:2023sfy}. Uncertainties in data
points are estimated using jackknife resampling with
$N_{\rm bin}=20$ bins. Uncertainties in splines use a 
parametric bootstrap with $N_{\rm boot}=500$ samples.
HRG model calculations use the particle list
QMHRG2020~\cite{Bollweg:2021vqf}.

\begin{figure}
\centering
\hspace{-5mm}
\includegraphics[width=0.5\linewidth]{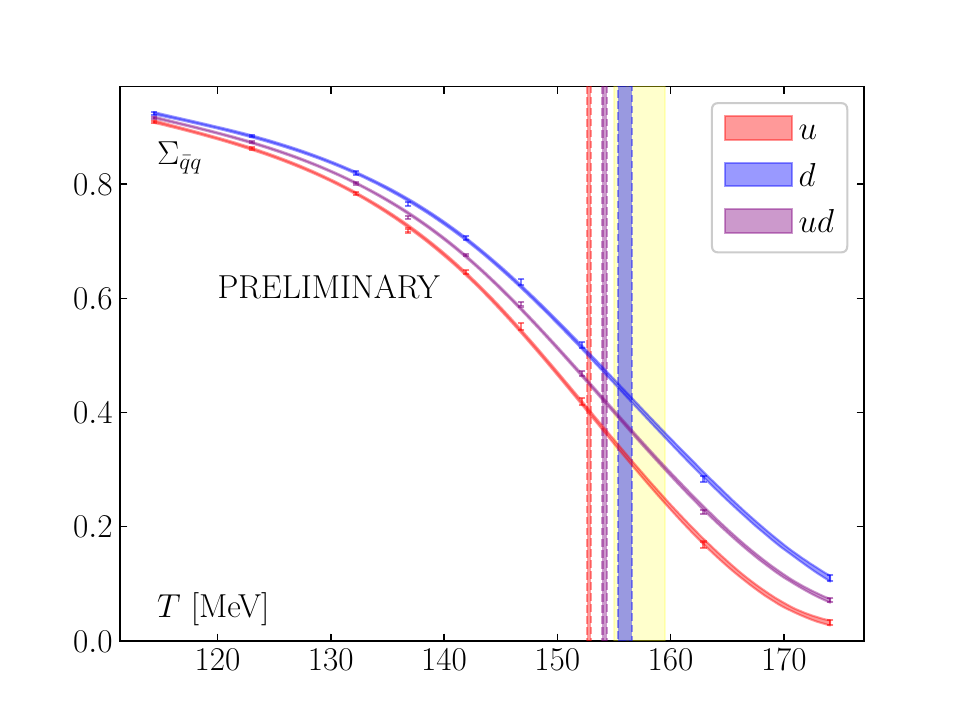}
\hspace{-3mm}
\includegraphics[width=0.5\linewidth]{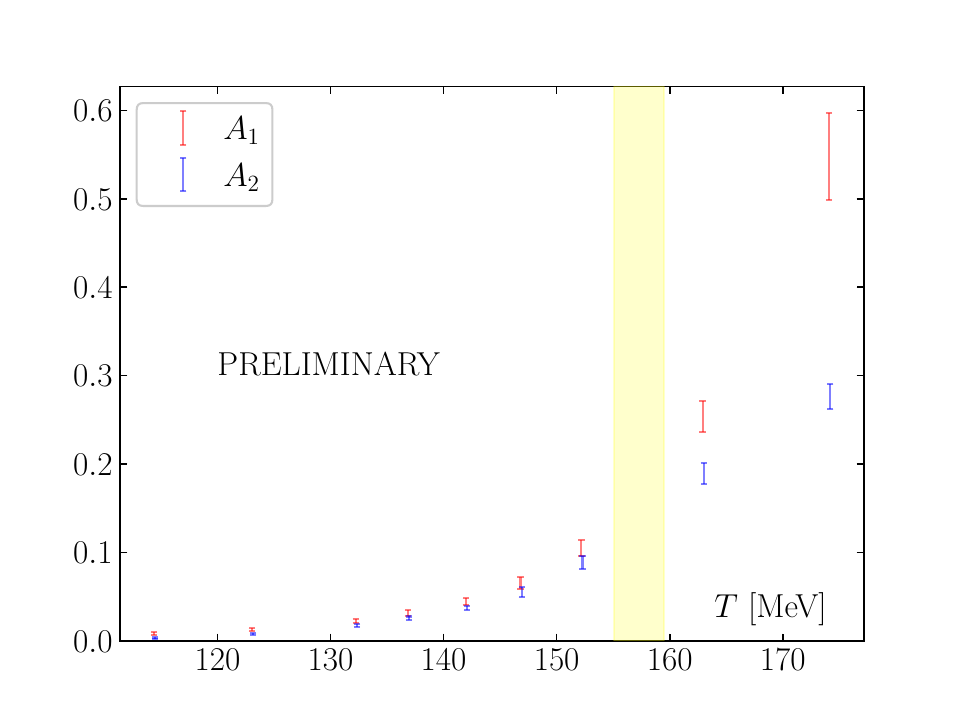}
\caption{Renormalized chiral condensates and abundances.
The yellow band indicates $\Tpc=157.3(2.3)$ MeV.
{\it Left}: Chiral condensates. Shown
are results at each temperature along with a spline fit.
Vertical, dashed bands indicate $\Tpc$ estimated from each
condensate. {\it Right}: Abundances $A_1$ and $A_2$. 
Data for $A_2$
are slightly displaced on the abscissa to improve visibility.}
\label{fig:renorm_condensate}
\end{figure}

\section{Results}

We begin with the chiral condensates and associated pseudocritical
temperatures, shown in \figref{fig:renorm_condensate} (left). 
The down-quark condensate is found to exceed the up-quark condensate
across the entire temperature range.
A pseudocritical
temperature can be defined by the inflection point in $\Sigma_{\bar q q}$. These estimates are also
shown in the figure. In particular we find for $N_\tau=8$ from the inflection point
in the renormalized condensate
    $\Tpc^{\Sigma,u}=152.79(20)$ MeV,
    $\Tpc^{\Sigma,ud}=154.18(20)$ MeV, and
    $\Tpc^{\Sigma,d}=155.98(60)$ MeV,
where uncertainties are estimated using a bootstrap
and no spline systematics are yet included.
We find that $\Tpc^{\Sigma,ud}$ just skirts the lower
edge of the crossover region, compatible within uncertainties,
suggesting that SIB effects are
negligible for the crossover temperature, in alignment with
Ref.~\cite{Gavai:2002fi}.

In \figref{fig:renorm_condensate} (right) we show the abundances $A_1$ and $A_2$.
The abundances increase monotonically with temperature
and largely agree below $\Tpc$, suggesting the $T=0$ limit
of $A_1$ does not appreciably tarnish its interpretation
in the explored temperature range. Both curves show a clear abundance
of $\ev{\bar d d}$ over $\ev{\bar u u}$ in concordance with
\figref{fig:renorm_condensate}. The curves indicate an abundance
of roughly 0.2 at the crossover. This may partially 
explain the $R_K$ puzzle; still, we emphasize that a quantitative
mapping from the condensate difference to $R_K$ is not yet clear.

\begin{figure}
\centering
\hspace{-5mm}
\includegraphics[width=0.45\linewidth]{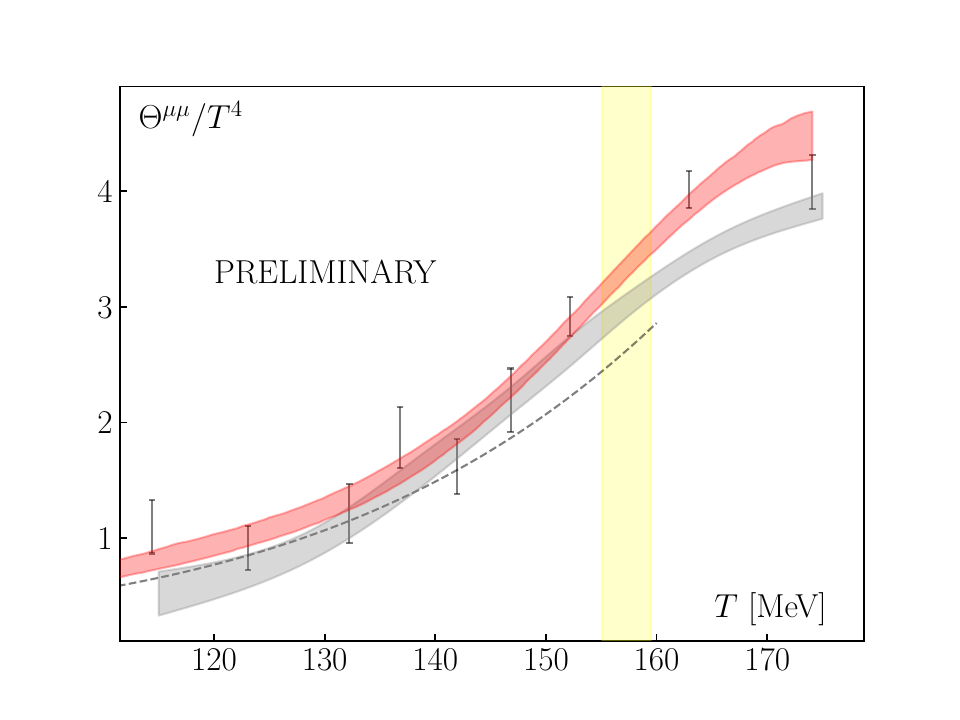}
\hspace{-3mm}
\includegraphics[width=0.45\linewidth]{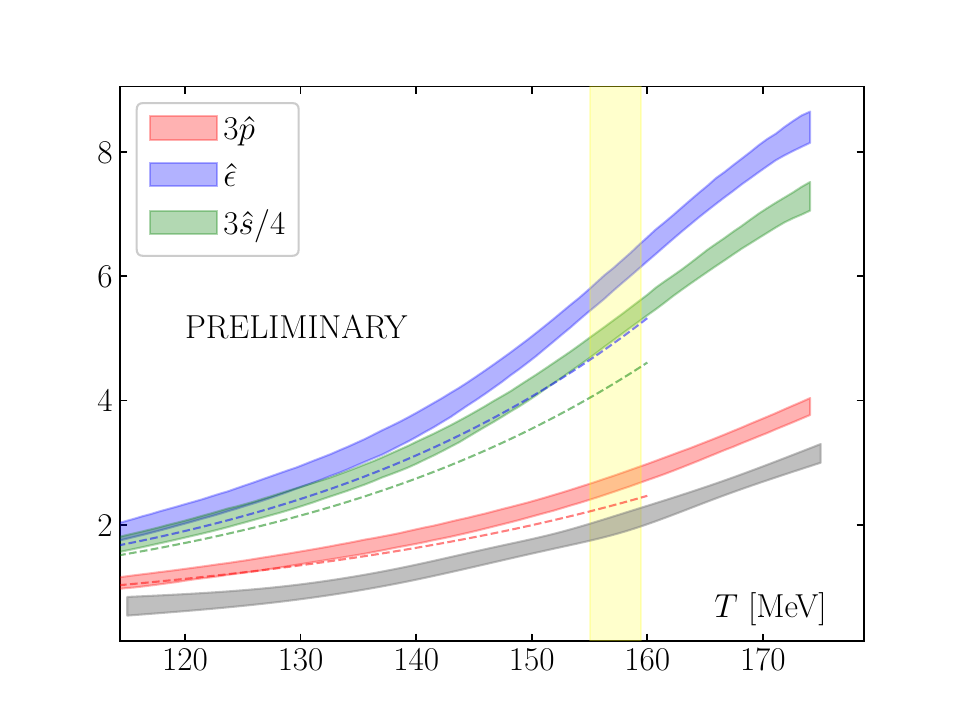}
\caption{QCD equation of state. The yellow band indicates $\Tpc$.
Dashed lines indicate HRG results.
Gray bands indicate
$N_f=2+1$, continuum-extrapolated results from Ref.~\cite{Borsanyi:2010cj}.
{\it Left}: 
The trace anomaly for $N_\tau=8$.
The red band shows a smoothing spline interpolation
of our $N_f=1+1+1$ data.
{\it Right}: The derived pressure, energy density, and entropy density.
}
\label{fig:traceAnomaly}
\end{figure}

In \figref{fig:traceAnomaly} (left) we show the trace anomaly.
For comparison, we plot as a gray band results from
Ref.~\cite{Borsanyi:2010cj} using the
same action but with degenerate light quark masses. We
generically see agreement
between the $N_f=1+1+1$ and $N_f=2+1$ results for all temperatures,
at least within our substantially larger uncertainties.
Computing the pressure requires an interpolation of this
quantity. The red band shows a smoothing spline fit.
We see small differences with
the $N_f=2+1$ continuum result for low temperatures, with a
larger discrepancy at high temperatures. These discrepancies
translate to differences in the pressure, which can be seen
in \figref{fig:traceAnomaly} (right). The difference between these pressures is further exacerbated
by different choices for $\Theta^{\mu\mu}(T_0)/T_0^4$ and $P(T_0)/T_0^4$:
QMHRG2020 delivers respectively for these quantities 0.422
and 0.273 compared to 0.41 and 0.16 from Ref.~\cite{Borsanyi:2010cj}; in particular if we adopt
their reference values at $m_\pi=135$ MeV for our data, we obtain a pressure curve that overlaps with Ref.~\cite{Borsanyi:2010cj} up to about 130 MeV before overtaking it. Therefore we currently cannot
attribute this difference to SIB effects.

\begin{figure}
\centering
\includegraphics[width=0.5\linewidth]{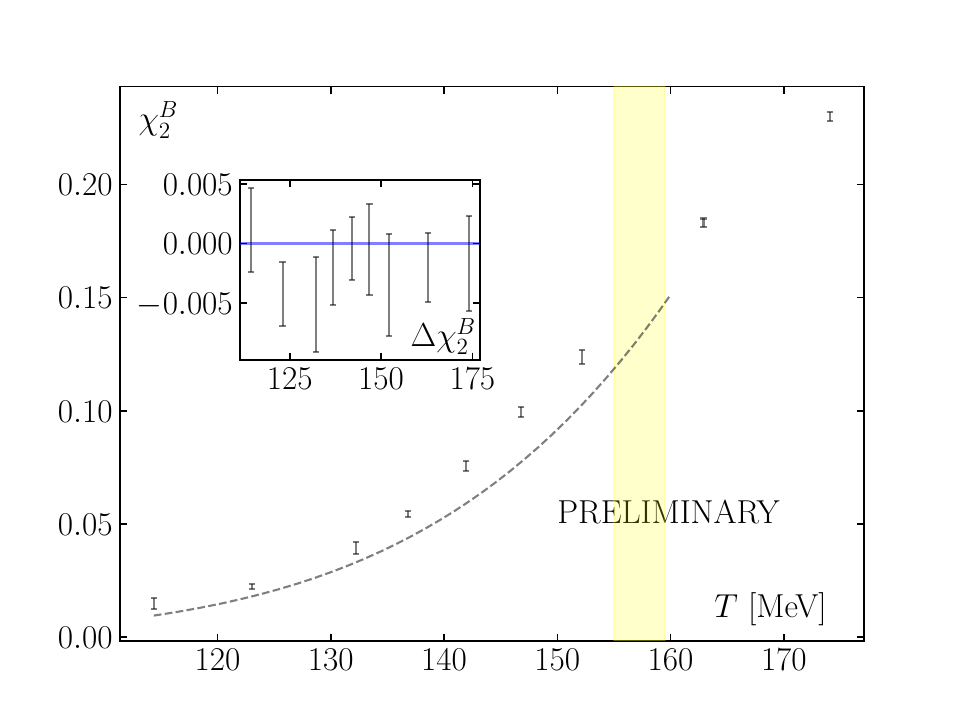}
\hspace{-3mm}
\includegraphics[width=0.5\linewidth]{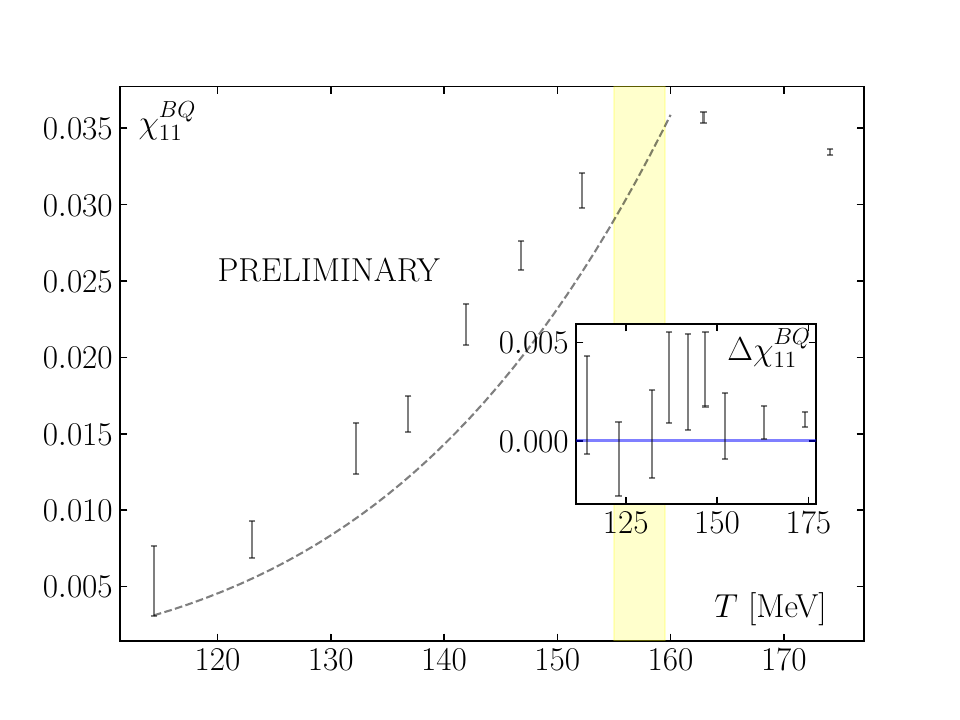}\\[-4mm]
\includegraphics[width=0.5\linewidth]{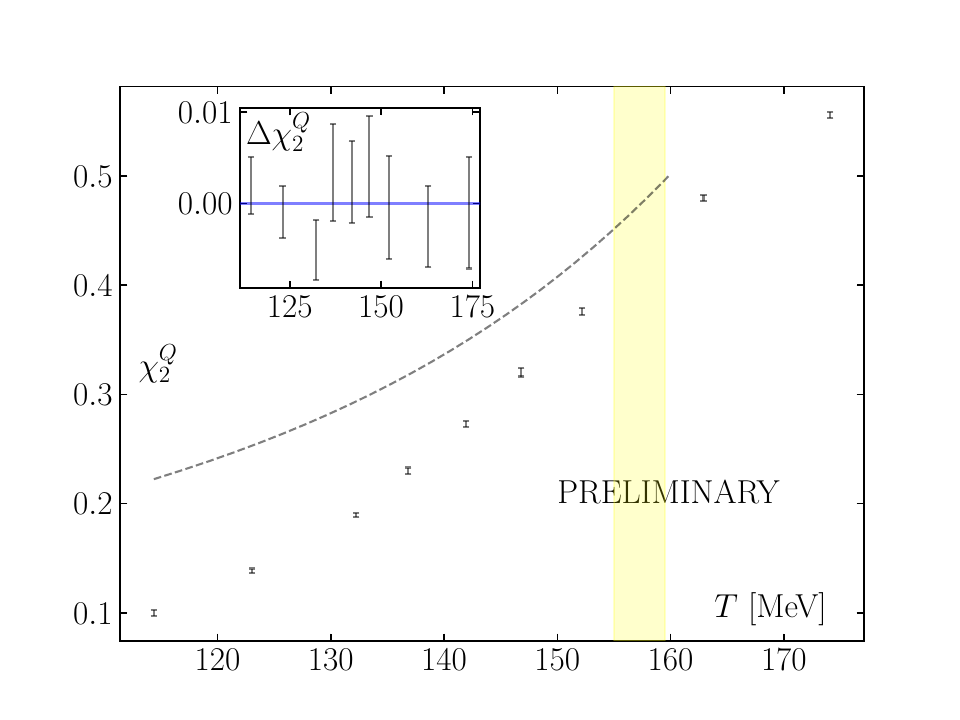}
\hspace{-3mm}
\includegraphics[width=0.5\linewidth]{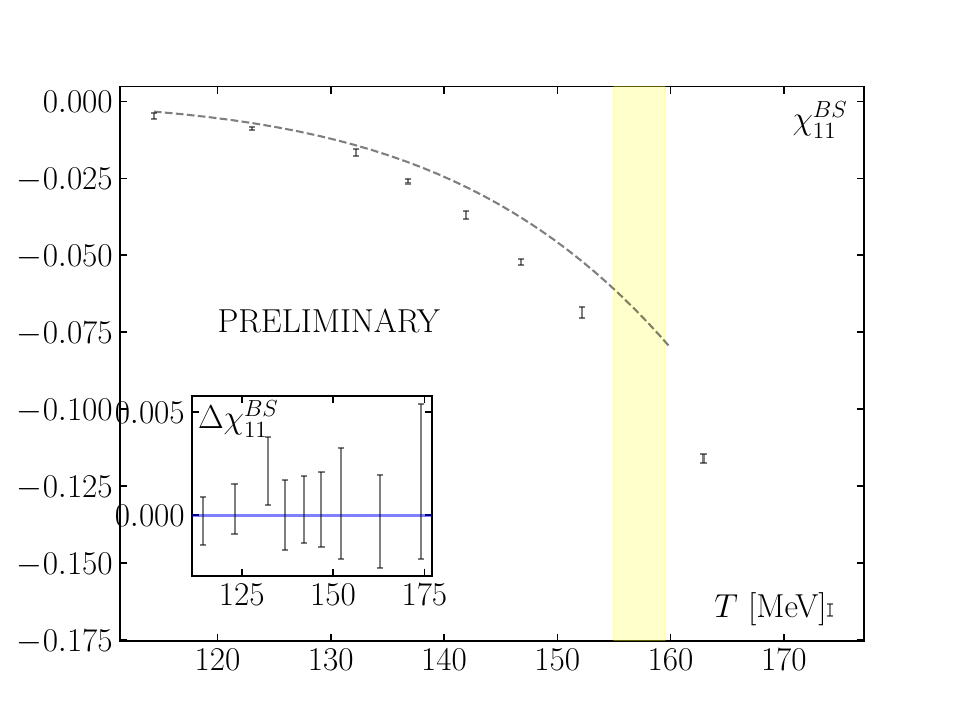}\\[-4mm]
\includegraphics[width=0.5\linewidth]{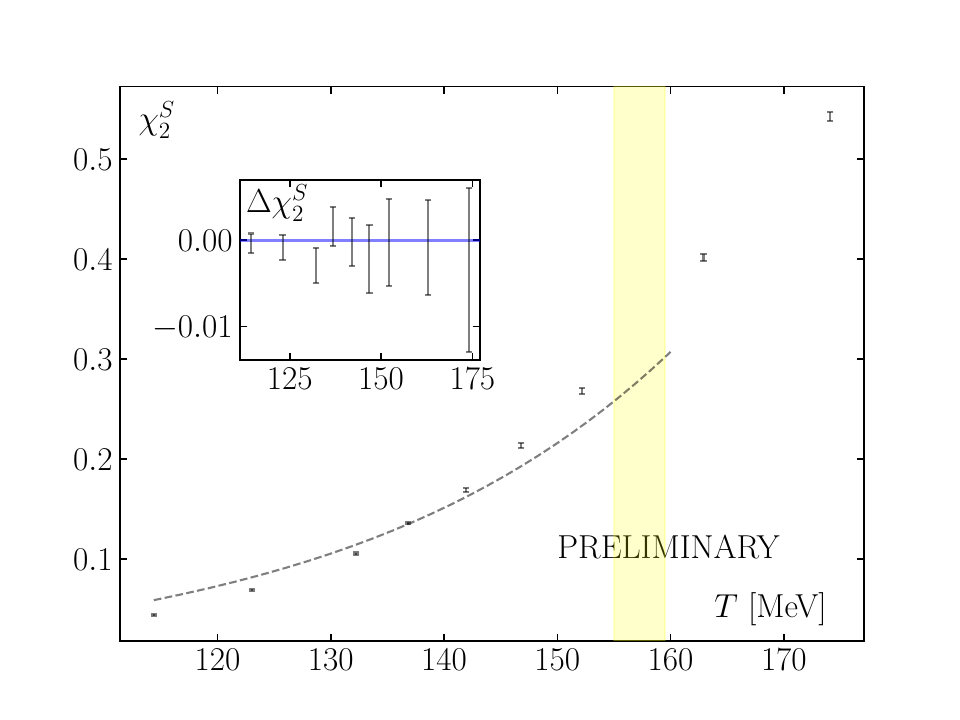}
\hspace{-3mm}
\includegraphics[width=0.5\linewidth]{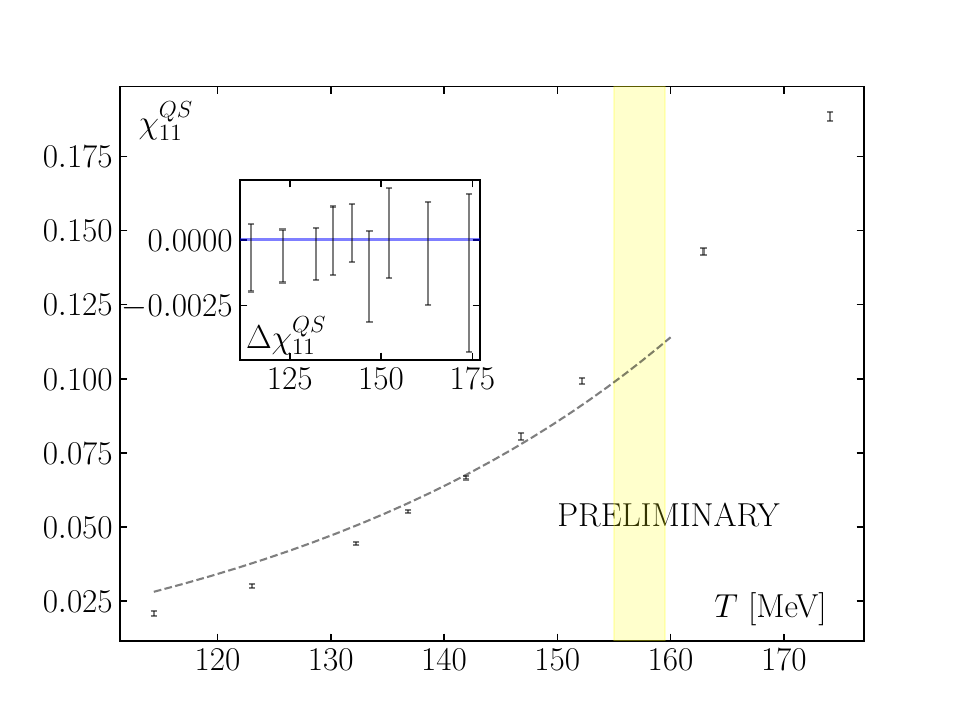}\\[-4mm]
\includegraphics[width=0.49\linewidth]{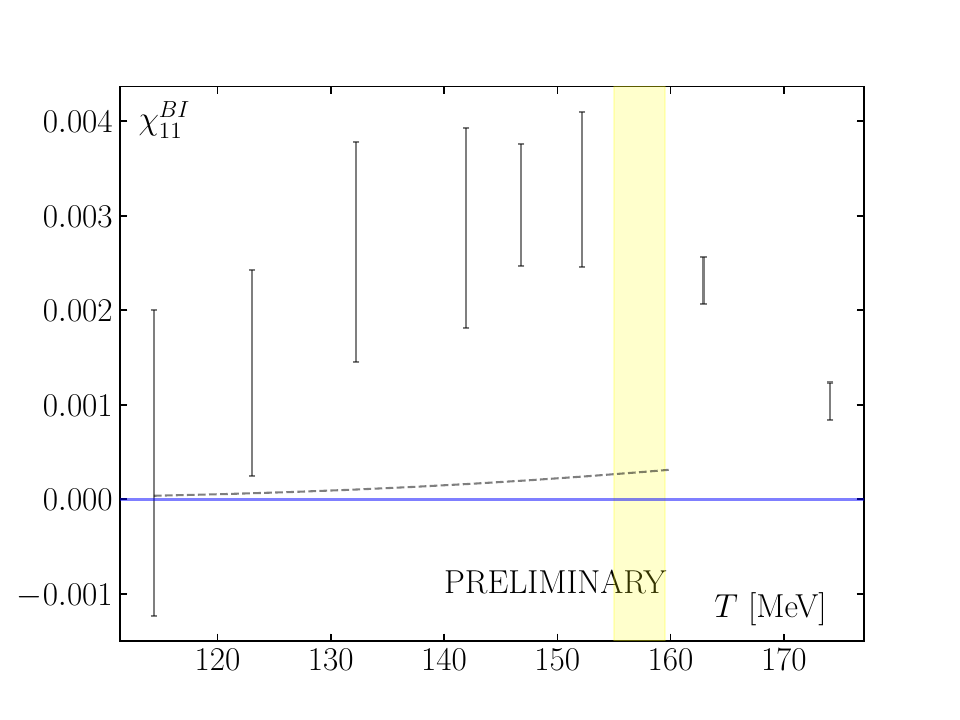}
\hspace{-3mm}
\includegraphics[width=0.49\linewidth]{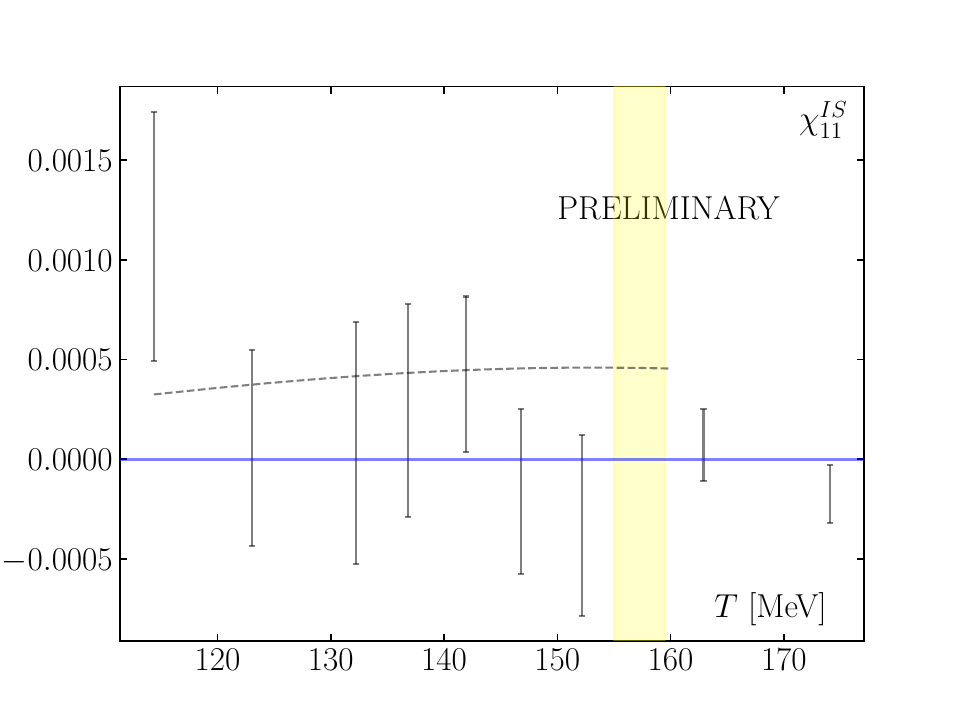}\\[-4mm]
\caption{Second-order, conserved-charge cumulants. The yellow band indicates $\Tpc$.
The dashed line indicates the HRG result.
Insets show the SIB correction.
}
\label{fig:2ndorder_cumulants}
\end{figure}

In \figref{fig:2ndorder_cumulants}, the first six plots 
show our second-order
cumulants in the $(B,Q,S)$ basis.
The insets show the
strong isospin-breaking effects; in particular we plot
$\Delta\chi\equiv\chi(m_u\neq m_d)-\chi(m_u=m_d)$,
where $\chi(m_u=m_d)$ is computed on our $N_\tau=8$,
$N_f=2+1$ ensembles. Normalizing 
$\Delta\chi$ to $\chi(m_u=m_d)$, we find SIB to 
be an at-most
3\% correction for all $(B,Q,S)$ cumulants
near $\Tpc$.

More interesting are off-diagonal isospin cumulants, shown in
the bottom two plots of \figref{fig:2ndorder_cumulants}. 
Here we set $\mu_I=\mu_u-\mu_d$ and show only
$N_f=1+1+1$ results, as these cumulants are zero in
the isosymmetric case. We see a clear signal for $\chi_{11}^{BI}$, 
and this cumulant
seems to peak below $\Tpc$.
This lines up qualitatively with a PNJL model calculation 
showing SIB leads to a peak below $\Tpc$ in this 
quantity~\cite{Bhattacharyya:2012up}.
The cumulant $\chi_{11}^{IS}$ shows however
no signal within our statistics.

\section{Summary}

In these proceedings, we showed our progress computing thermodynamic
observables in $N_f=1+1+1$ QCD at $N_\tau=8$. 
Currently most observables exhibit
at-most small effects, which are mostly negligible in practice.
In particular, the isosinglet
$\Tpc$ agrees up to uncertainty with the literature result,
in agreement with the findings of Ref.~\cite{Gavai:2002fi}.
Both renormalized abundances show a clear excess of
$\ev{\bar d d}$ over $\ev{\bar u u}$, which may have implications
for the unexpected abundance of charged kaons compared to neutral
kaons observed in collision experiments. Finally, we find a
distinct signal for $\chi_{11}^{BI}$ on lattices with broken isospin,
qualitatively in agreement with PNJL~\cite{Bhattacharyya:2012up}.
We are currently increasing statistics, refining our analysis,
and extending our analysis to more thermodynamic observables.

\acknowledgments

This research has received funding from the programme "Netzwerke 2021",
an initiative of the Ministry of Culture and Science of the
State of Northrhine Westphalia.
We thank Gergely Endrődi for discussions and help 
with the \texttt{dyniso} code. We also
acknowledge discussions with Joe Kapusta, Rob Pisarski,
Lorenz von Smekal and the participants of the ISO-BREAK 25
workshop.

\bibliographystyle{JHEP}
\bibliography{bibliography}

\end{document}